\documentclass{article}

\usepackage{microtype}
\usepackage{graphicx}
\usepackage{multirow}
\usepackage{minted_copy}

\makeatletter
\let\@float@c@listing\@caption
\makeatother

\newcommand{\il}[1]{{\fontfamily{qcr}\selectfont #1}}

\newcommand{\IMPROVPYTORCH}{74\% }
\newcommand{\IMPROVPYTORCHCPU}{38\% }
\newcommand{\IMPROVGIL}{33\% }

\usepackage{hyperref}

\usepackage[accepted]{mlsys2025/mlsys2024}

\mlsystitlerunning{Scalable and Performant Data Loading}

\begin{document}

\twocolumn[
\mlsystitle{Scalable and Performant Data Loading}



\mlsyssetsymbol{equal}{*}

\begin{mlsysauthorlist}
\mlsysauthor{Moto Hira}{mt}
\mlsysauthor{Christian Puhrsch}{mt}
\mlsysauthor{Valentin Andrei}{mt}
\mlsysauthor{Roman Malinovskyy}{mt}
\mlsysauthor{Gael Le Lan}{mt}
\mlsysauthor{Abhinandan Krishnan}{mt}
\mlsysauthor{Joseph Cummings}{mt}
\mlsysauthor{Victor Bourgin}{mt}
\mlsysauthor{Olga Gerasimova}{mt}
\mlsysauthor{Miguel Martin}{mt}
\mlsysauthor{Gokul Gunasekaran}{mt}
\mlsysauthor{Yuta Inoue}{mt}
\mlsysauthor{Alex J Turner}{mt}
\mlsysauthor{Raghuraman Krishnamoorthi}{mt}
\end{mlsysauthorlist}

\mlsysaffiliation{mt}{Meta Platforms, Inc, Menlo Park, California, USA}

\mlsyscorrespondingauthor{Moto Hira}{moto@meta.com}

\mlsyskeywords{Machine Learning, MLSys, AI, Data Loading, PyTorch, SPDL, DALI}

\vskip 0.3in

\begin{abstract}
We present SPDL (Scalable and Performant Data Loading), an open-source, framework-agnostic library designed for efficiently loading array data to GPU.
Data loading is often a bottleneck in AI applications, and is challenging to optimize because it requires coordination of network calls, CPU-bound tasks, and GPU device
 transfer.
On top of that, Python's GIL (Global Interpreter Lock) makes it difficult to gain performance improvement from multi-threading.
We found that when data preprocessing functions release the GIL entirely, it is possible to execute them concurrently in a thread pool, thereby improving the workflow performance.
Our benchmark shows that compared to the PyTorch DataLoader, SPDL can iterate through the ImageNet dataset \IMPROVPYTORCH faster while using \IMPROVPYTORCHCPU less CPU and 50GB less memory.
When training ViT-B/16 model, SPDL can send data to the GPU at a speed that does not starve the training.
Additionally, when using SPDL on Python 3.13t, without changing any code, the throughput is further by improved by \IMPROVGIL, thanks to the disabled GIL.
SPDL can improve the performance of current AI model training, and receives further performance improvements when Free-Threaded Python is adopted in production systems.
SPDL is available at \url{https://github.com/facebookresearch/spdl}.

\end{abstract}
]

\printAffiliationsAndNotice{}

\section{Introduction}

Training ML models takes time and consumes significant computational resources. GPUs can help accelerate the computation, but feeding data to GPUs continuously is challenging.
As GPUs become faster, data needs to be transferred to GPUs faster. And as models become larger, more data is required to improve the quality of the models. \cite{kaplan2020scaling}
Data loading is one of the most common bottlenecks in model training.
This is because efficient data loading requires coordination across data acquisition, pre-processing and data transfer to the GPU.
Each of these steps have different bounding factors.
Table \ref{table:bounds} summarizes this.

\begin{table}[t]
\caption{Typical data loading stages, factors bounding the performance of each stage, and approaches ML practitioners can take to make the data loading more efficient.}
\vskip -0.07in
\label{table:bounds}
\begin{center}
\begin{small}
\begin{tabular}{p{0.25\linewidth}p{0.22\linewidth}p{0.37\linewidth}}
\hline
Stage
  & Factors
  & Approaches
  \\ \hline
Data acquisition
  & Network Latency
  & Make multiple network calls concurrently
  \\
\cline{2-3} 
  & API Rate limit
  & Switch to a service that has a higher rate limit
  \\ \hline
\multirow{2}{*}{Pre-processing} 
  & Memory
  & Avoid unnecessary memory copies
  \\
\cline{2-3} 
  & CPU
  & Process multiple data concurrently.
  \\
  & 
  & Use disaggregated systems on multiple hosts
  \\
  \hline
GPU transfer
  & PCIe bandwidth
  & Continuously transfer data to GPU without interrupting the kernel launches.
  \\
  \hline
\end{tabular}
\end{small}
\end{center}
\vskip -0.33in
\end{table}

In addition, Python's GIL prevents the effective use of multi-threading. 
As a workaround, many data loading solutions resort to multi-processing, which provides some performance boost.
However, multi-processing comes with drawbacks due to it performing operations in isolated memory and then transferring resulting data to the main process. 
Section \ref{section:subprocess} describes these drawbacks in detail.

Currently, there are huge efforts to remove the GIL from Python \cite{pep703} which would allow Python programs to utilize multi-threading.
Switching from multi-processing to multi-threading eliminates the overhead of inter-process communication.
Once the free-threaded (FT) Python becomes production-ready, data loading can enjoy the benefits and its performance should improve.
There are high hopes that FT-Python makes data loading faster, but questions arise.
\begin{enumerate}
    \item How much of a performance boost is possible?
    \item What additional steps are required to achieve the performance boost?
    \item Is it possible to bring performance benefits to the current Python versions (3.12 and below)?
\end{enumerate}

To answer these questions, we experimented with thread-based parallelism, using regular Python, approximating the effect of FT-Python with releasing the GIL.
If functions release the GIL, they can run concurrently in a thread pool.
If we structure the parallelism of data loading in a way that minimizes the competition for the GIL, then we can achieve a throughput higher than data loading solutions based on process-based parallelism.
Applying these techniques, SPDL achieved end-to-end throughput \IMPROVPYTORCH higher than the PyTorch DataLoader \cite{pytorch} on the Image Classification task using ViT-B/16 \cite{vit} and the ImageNet dataset \cite{deng2009imagenet}.
We also verified that the data loading pipeline written with SPDL works with FT-Python, and without changing the code, its performance was further improved by \IMPROVGIL compared to Python 3.12.

The rest of this paper is composed as follows.
Section \ref{section:bound} goes over common challenges encountered in data loading.
Section \ref{section:subprocess} looks at the drawbacks of process-based parallelism.
Section \ref{section:process_vs_thread} shows experimentally that releasing the GIL allows to scale the performance in multi-threading. 
Section \ref{section:design} describes the design of the proposed SPDL library.
Section \ref{section:benchmark} shows the benchmark against popular, publicly-accessible data loading solutions.
Section \ref{section:ft} compares the performance between regular Python and FT-Python to estimate how much performance gain SPDL is realizing in regular Python.

\section{Overview of data loading} \label{section:bound}
\subsection{Stages and bounding factors}

As previously mentioned, data loading is challenging because it is composed of multiple stages bounded by different factors. See table \ref{table:bounds}.

The first stage is data acquisition.
When the volume of the training data is large, it needs to be retrieved from remote storage.
This process is bounded by network bandwidth, or an API rate limit on the remote storage system.
These are restrictions imposed by the environment, thus not changeable by ML practitioners.
But within these limitations, making multiple fetch requests concurrently can improve the overall throughput.

Once the data is retrieved, it must be pre-processed before it can be transferred to GPUs. Files in remote storage are often encoded for storage efficiency and faster network transfer. For example, multimedia data are stored in formats such as JPEG, FLAC and MP4. Therefore, the first step of pre-processing must be decoding. After decoding, the data is further converted to a form suitable for batching. This often includes resizing image frames, or resampling audio. Frequently, decoding and pre-processing media data show up as bottlenecks in data loading. Video decoding is especially CPU-intensive, but operations like resizing and resampling many samples can also consume substantial CPU resources.

Batching is the process of combining multiple pre-processed data samples to take advantage of a GPUs ability to process data quickly via matrices. It requires allocating one contiguous memory for all the samples in a batch. This can be a non-trivial amount of contiguous memory. For example, a batch of 32 RGB images with resolution of 224 pixels in width and height occupies about 4.8 mega bytes. It is ideal to avoid re-allocating such memory spaces repeatedly.

Finally, the batch is sent to GPUs. The data must be sent to GPUs in a way such that the model computation and data transfer do not interrupt each other. To achieve this, in addition to transferring data in the background, the data must be prepared in page-locked memory and transferred to GPU via a stream different from the one used for model computation. Due to hardware limitations, it is not possible to transfer multiple chunks of contiguous data to GPUs concurrently. Therefore, there should be at most one task handling the transfer stage.

\subsection{Parallelization}

Since training samples are generally independent of each other, it is conceptually simple to split the work of pre-processing into independent jobs and execute them concurrently.
As we described before, we can make multiple fetch requests to obtain the source data from remote storage. We can also create multiple decoding jobs to process samples concurrently.

Additionally, we can run all the stages concurrently too. Such an approach is often referred to as software pipelining \cite{pipelining} and also applied to distributed systems \cite{pipeline_training} \cite{gpipe}. However, as noted before, the GIL makes it difficult to take advantage of more elaborate parallelization schemes.
A typical workaround is to use process-based parallelism and run the entire data loading pipeline in each process.

\section{Issues with multi-processing} \label{section:subprocess}

Multi-processing gives a performance boost, but it comes with drawbacks. Here we summarize the common issues encountered in process-based data loading.

\subsection*{Launch time and static memory consumption}
When a subprocess worker is launched, a Python interpreter is created and libraries are initialized. \cite{mp} Some libraries take non-negligible time to initialize. This gets exacerbated when the number of workers is increased, as the main process launches subprocess workers sequentially.



\subsection*{Inter-process communication (IPC)}
Data exchanged across the boundary of processes are serialized into a byte string first, then written to the memory region shared between processes.
The recipient process de-serializes the data from the byte string. Typical data exchanged between processes are a list of data source information (commonly referred to as a Dataset), and sample batches created in the subprocess.
When a Dataset instance contains a large list of source information, copying the instance to the worker process incurs significant time. This is shown in table \ref{tbl:ttfb}.

For exchanging array data, some libraries have custom implementations.
For example, PyTorch writes the array data into shared memory in a way that the main process can directly reference it without making an extra copy.

\subsection*{Sequential serialization in IPC}
Even when using subprocesses, the Python interpreter in the main process is still constrained by the GIL. Therefore, when the main process retrieves data created by subprocesses, the main process must serialize them one-by-one, and it cannot perform other operations meanwhile.

\subsection*{Inability to synchronize objects}
When using multi-processing, data can be exchanged through IPC; however, their state cannot be synchronized in such a manner. 
This makes it challenging to sample from a dataset without overlaps and keep track of data that are sampled to prepare for the case where training is halted and resumed.

\section{Releasing the GIL in multi-threading} \label{section:process_vs_thread}

In this section, we examine the scalability of functions that release the GIL and leverage concurrency through multi-threading.

Functions defined in extension modules can call Python C APIs for releasing and re-acquiring the GIL without additional overhead.
Libraries like NumPy \cite{numpy} and PyTorch release the GIL when it dispatches the operation to low-level numerical computation routines.

To understand the scalability of preprocessing functions, we implement batch image loading functions and run them in a thread pool and a process pool, comparing the throughput.
For the baseline, we used the Pillow library \cite{clark2015pillow}, which releases the GIL partially but not when decoding and resizing the image.
As a comparison we implemented the same functionality with SPDL's IO module.
The detail of SPDL's design is given in Section \ref{section:design}, but the key aspect here is that SPDL's IO module releases the GIL entirely.
We also ran this implementation on FT-Python to see its scalability and performance improvement.

\begin{figure}[t]
\begin{center}
\caption{The peak throughput of media processing functions in multi-threading and multi-processing. Images are decoded, resized and converted to RGB, then batched. The batch size is 32. \il{ThreadPooolExecutor} and \il{ProcessPoolExecutor} from \il{concurrent.futures} standard module are used. The test environment has 96 CPU cores. The time to initialize the workers, which is non-trivial for process workers, is not included.}
\vskip -0.12in
\label{fig:process_vs_thread}
\includegraphics{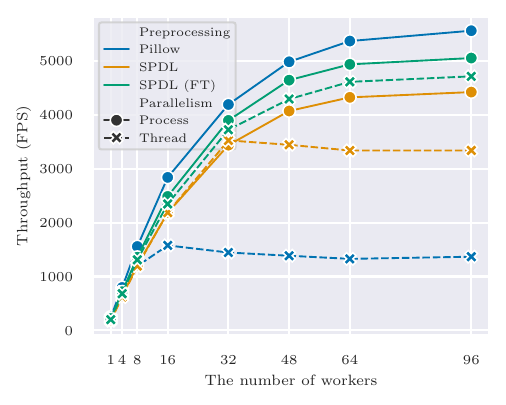}
\end{center}
\vskip -0.3in
\end{figure}

Fig. \ref{fig:process_vs_thread} shows the throughput of the resulting data loading pipelines.
The experiment was conducted on an AWS \il{p4d.24xlarge} instance, which has 96 CPU cores. \cite{p4d}
The code and the detail is found in Appendix \ref{appendix:process_vs_thread}.
Note that for the sake of simple comparison, the measurement does not include the time it took to initialize the workers, which can be non-trivial for multi-processing as we will see in Section \ref{section:benchmark}.
We can see that all pipelines scale with multi-processing.
Their performance improvement gets saturated at the number of CPU cores.
In multi-threading, Pillow achieves the peak performance at 16 threads and its performance stagnates above that.
SPDL with regular Python is similar but scales up to 32 threads. 
SPDL with FT-Python, on the other hand, shows a similar scaling trend as multi-processing.

\begin{figure}[t]
\caption{The time primitive operations take to execute. \il{open} is Python's built-in function and \il{ImagingCore} and \il{ImagingDecoder} are Pillow's low-level functions. The 0 concurrency means the image loading is executed in the main thread. Other concurrency values are the number of threads.}
\includegraphics{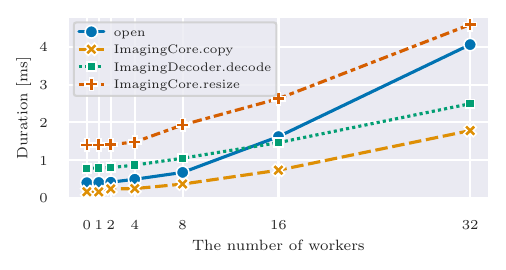}
\label{fig:thread_traces}
\vskip -0.4in
\end{figure}

When process workers perform more operations, there are more operations executed concurrently across these subprocesses.
On the contrary, when using thread pool, even though some parts of the pipeline release the GIL and are executed concurrently, other parts still compete to acquire the GIL.
So increasing the number of threads increases the number of contenders for the GIL, which has a negative impact on the completion of data loading operations.
Fig. \ref{fig:thread_traces} shows the average execution time of some of the functions of Pillow pipeline in the thread pool.
We can see that as the number of threads are increased, the duration of these functions grows.

The throughput of SPDL's IO module peaks at 32 threads, and does not scale all the way to the 96 cores.
In practice, this is good enough because the \il{p4d.24xlarge} instances are equipped with 8 GPUs, and one main process is created for each GPU.
This means that at most 12 concurrency can be used by one process.
The scalability of SPDL's IO module is good enough to utilize all the CPU cores in actual training.

\section{Designing scalable data loading} \label{section:design}

\subsection{Alternative parallelization for multi-threading} \label{section:alt-parallel}
The observation from Section \ref{section:process_vs_thread} suggests that it is possible to scale the performance with multi-threading when the GIL is released entirely. However, simply replacing process-based parallelism with thread-based parallelism does not improve the throughput, because data loading pipelines involve parts that do not release the GIL.

This gives us an idea that if we can rearrange the parallelism in a way that restricts the GIL competition to a small number of threads, and if we use the thread pool for running functions that release GIL only, we might be able to scale the performance of overall data loading pipelines with multi-threading.

\begin{figure}[t]
\caption{Illustration of the GIL contention in multi-threading and different approaches for parallelization. Left: Parallelizing the entire data loading operation. The main thread and all the worker threads compete for the GIL, resulting in an ineffective use of the threads. Right: Parallelizing primitive operations separately. The scheduler thread dispatches operations that release the GIL to a thread pool. This way, the threads are not blocked. The competition to acquire the GIL is limited between the main thread and the scheduler thread. This approach also allows to configure the parallelism for different steps.}
\includegraphics{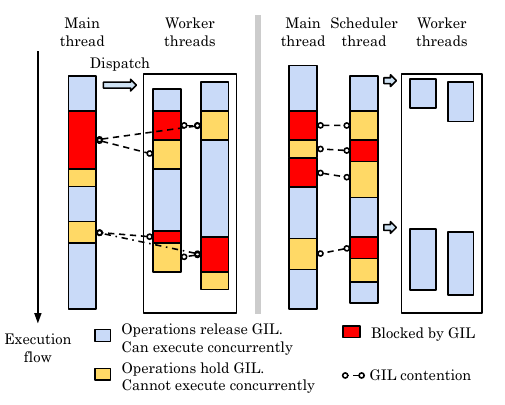}
\vskip -0.2in
\label{fig:re-architect}

\end{figure}

Many data loading solutions, especially those that adopt process-based parallelism, concurrently execute the entire pipeline.
An alternative approach is to parallelize parts of the pipeline separately.
For example, we can run only the performance-critical code (such as data acquisition and media processing) in a thread pool, and let other operations run in a separate dedicated thread.
We expect that by limiting the functions executed in a thread pool to those that release the GIL, there is less competition for acquiring the GIL amongst the other functions, therefore, we can utilize the thread pool more effectively, and achieve better higher overall throughput.
For this reason, we introduce a scheduler thread with the intention that the competition for the GIL will be mostly limited to the main thread and the scheduler thread. 
The responsibility of the scheduler thread is to dispatch the parts of data loading pipeline to the thread pool, and handle its completion. 
See Figure \ref{fig:re-architect}.

\subsection{Support for Coroutines}

So far, we have discussed the performance of media processing in the context of multi-threading vs. multi-processing.
Now we turn our attention to another critical component of data loading, which is data acquisition.

Generally speaking, network calls are the slowest kind of operations in computing. To facilitate concurrent networking, Python supports asynchronous execution of coroutines. Many networking libraries provide asynchronous interfaces to take advantage of this.
One important aspect of coroutine functions is that they are not constrained by the GIL.
Some implementations use non-blocking functions provided by the operating system, and other implementations resort to multi-threading in low-level languages like C/C++.
Coroutine functions by nature can be executed concurrently without interrupting other Python function calls with the GIL.\footnotemark{}
Therefore, when building high-performance data loading pipelines, it is important to take advantage of coroutine functions and incorporate their asynchronous execution model.

\footnotetext{It is possible to convert a regular function into coroutine function by the help of threading and an async event loop. However such coroutine functions are still constrained by the GIL.}

\subsection{Learning from existing solutions}

User experience is also important when designing solutions for ML practitioners. Here we look at popular data loading libraries and take hints for improved usability.

\subsubsection{PyTorch DataLoader}

PyTorch separates data loading into \il{Dataset} and \il{DataLoader}  \cite{pt_dl}. \il{Dataset} defines a mapping from key to data. This logical separation is easy to adopt but it hides the internal mechanism of loading data from the \il{DataLoader}.
As a result PyTorch DataLoader provides only one option for optimizing performance: number of worker processes.

The separation of \il{Dataset} and \il{DataLoader} encourages the creation of intermediate tensors.
For example, when an image is decoded, its color planes are usually separately allocated and padded for better memory alignment.
The \il{Dataset} interface returning tensor means that it converts this internal representation into a tensor, which is discarded once the batch tensor is created.
It is more efficient if the data are copied from the internal representation to batch tensor only once, but the \il{Dataset} interface discourages such design.

\subsubsection{DALI}

DALI \cite{dali} employs thread-based parallelism by converting the Python code into its own execution format. DALI extensively uses a DSL (Domain-specific language) to define the pipeline description and the custom operations supported by the pipeline.

The DSL defines the subset of language features that users can use in the pipeline and requires users to learn its semantics.
When the DSL is sufficiently elaborate, it becomes difficult for users to learn.
The maintenance cost of the library itself increases because the combination of DSL features that need to be tested grows exponentially as new features are added.

\subsubsection{FFCV}

FFCV \cite{leclerc2023ffcv} is a data loading library for computer vision tasks. It employs a proprietary file format to make the runtime fast. This approach works for well-established tasks and datasets, but it lacks flexibility.
FFCV requires the dataset to be converted to its own format first. This additional step must be incorporated into the model training pipeline which in turn adds complexity to the infrastructure and its maintenance, and possibly increases storage costs by having to store the converted datasets as well.
In model development, we often combine multiple datasets in attempt to improve the model quality.
FFCV's custom dataset and data loader solution do not support this out-of-box.
Finally, since the data format is governed by FFCV, it's difficult to make modification to the format. 

\subsubsection{Decord} \label{section:decord}

Decord \cite{decord} is a library for loading videos.
Decord's \il{VideoLoader} class is designed to operate on a mapping from index to a batch across video files. To create this mapping, \il{VideoLoader} opens and queries all the video files sequentially at initialization. The time it takes for \il{VideoLoader} to operate scales with the size of the dataset. In addition, it is not robust against failure. When a video file is malformed (which happens quite often), \il{VideoLoader} fails to initialize and does not proceed.
On top of this, \il{VideoLoader} keeps all the decoder objects from initialization time for background processing. This means that the amount of compute resource (such as memory and file descriptors) is not bounded.

\subsection{Design Principles} \label{principle}

With the direction of new parallelization and learning from exiting solutions, we list below the design principles of SPDL.

\begin{description}
\item[High throughput]
The most important problem that must be solved is performance. So as to keep GPUs busy with computation, the data loading pipeline must achieve high-throughput.
\item[Visibility]
The data loading consists of multiple stages with different bounding factors. To optimize the performance, we need to be able to tell which stage is the bottleneck and for what reason.
\item[Tunability]
Once we identify the bottleneck in the data loading pipeline, we need to be able to tweak the relevant stages. Separating the data loading into Dataset and DataLoader and encapsulating the pipeline logic into Dataset prevents this.
\item[No domain-specific language]
The onboarding cost for using the new data loading should be as small as possible so that ML practitioners can focus on model training. This discourages the introduction of DSL. Also maintaining and extending DSL requires a lot of work, and the amount of effort grows exponentially as the DSL grows.
\item[Seamless asynchronous support]
We assume data are retrieved from network storage. Many network utilities provide asynchronous API for better performance. Asynchronous operations are not constrained by the GIL, therefore, the pipeline should support performing asynchronous methods seamlessly.
\item[Flexibility]
The way data are prepared varies from pipeline to pipeline. Sometimes data are in an archive format, so the whole archive has to be downloaded. Some pipelines have to gather data from different sources. To support these different needs, data loading framework must be able to build wide variety of pipelines.
\item[Robustness]
The pipeline needs to be robust against sample processing failures. 
For example, data acquisition can fail due to a failure in network calls or an interruption of remote service.
Decoding media data can fail due to malformed data.
The pipeline needs to log such failures and signal the errors to pipeline owners.
\item[Framework-agnostic]
Data loading is a common challenge in ML. Moving data from remote storage to GPU can be decoupled from how the data are consumed afterwords meaning the data loading library can be independent from any one deep learning library.
\end{description}

\subsection{Architecture}

Following the design principles listed in Section \ref{principle}, we designed the core of SPDL in the following manner.
Fig. \ref{fig:spdl-architecture} illustrates the architecture of SPDL.

\subsubsection{Task Scheduler}
    We use an asynchronous event loop as a core task scheduler. Asynchronous event loops schedule tasks (coroutine executions), wait and react to their completion. The event loop also supports running synchronous functions by delegating the execution to a single thread. This allows SPDL to support both synchronous and asynchronous functions, mixing them seamlessly. An event loop has a default \il{ThreadPoolExecutor} object, so the asynchronous event loop can also manage the life cycle of the thread pool.
\subsubsection{Scheduler thread and thread pool}
    Since running the asynchronous event loop itself is a blocking operation, it cannot be executed in the main thread. So we spawn a thread dedicated for running the event loop and let the event loop manage the job dispatch to the worker thread pool. This composition also aligns with the approach for less GIL contention depicted in Fig. \ref{fig:re-architect}.
    For multi-threading, we use the \il{ThreadPoolExecutor} attached to the asynchronous event loop as-is. New threads are created when the event loop dispatches the execution of synchronous functions to the thread pool and the thread pool does not have an available thread. 
\subsubsection{Queues}
    The stages of the data loading pipeline are connected with queues. A task belonging to the stage gets one item from the input queue and puts the result in an output queue. If the output queue is full, then the task is blocked until a slot becomes available. This structure propagates the congestion from downstream (model training happening in the main thread) towards upstream. It responds quickly when the congestion is resolved.
    If the downstream is slow, then the queue at the sink gets full, and tasks start to get blocked. The state of the congestion bubbles up to the source, and eventually the whole data loading pipeline gets blocked.
    Once the main process fetches one item from the sink, then a task being blocked on putting an item to the sink will complete, and the event loop will schedule the next task.
    This will free-up a slot in the queue one upstream to the sink, so one task being blocked on the queue will now complete.
\subsection{Stage definitions} \label{section:stage}
    Each stage is defined by a function. SPDL does not dictate what operations go into the function, what input should be passed and what output is expected. The stage function must meet the requirement of releasing the GIL and thread-safety, but other than that, it entirely depends on users how stages are defined. We have discussed that data loading consists of data acquisition, pre-processing, and device transfer, but SPDL does not require such compositions. Users can breakdown processes in any ways they want.

\begin{figure}[t]
\caption{The architecture of SPDL. An asynchronous event loop runs in a background thread. This event loop schedules tasks of different stages (data acquisition, processing and device transfer), and handles their completion. Asynchronous functions are executed as-is. They do not necessarily use the thread pool. Synchronous functions are delegated to the thread pool, so that they are handled in the same manner as native asynchronous functions. The data handle (Tensor/Array objects) are passed to the main thread via queue.}
\vskip - 0.1in
\includegraphics{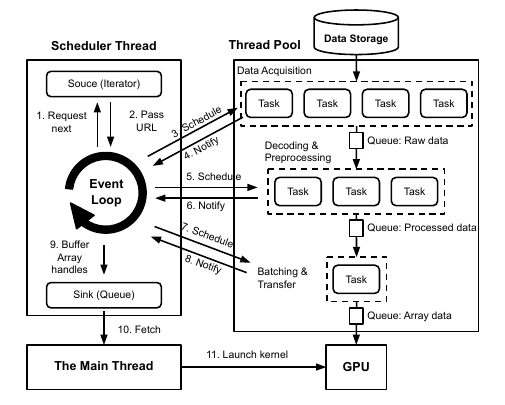}
\label{fig:spdl-architecture}
\vskip -0.3in
\end{figure}

Fig. \ref{fig:spdl-architecture} illustrates the architecture of SPDL. The data loading engine depends only on Python standard libraries. Media processing functions and GPU transfer operators are written in C/C++ and they release the GIL. Network utilities for data acquisition are expected to be provided by third parties. 

\subsection{Supplemental components}

As mentioned in \ref{section:stage}, users can use any function to build the data loading pipeline. However SPDL also provides high-performance functions for media processing and GPU data transfer. 

\subsubsection{I/O functions}
    To take best advantage of multi-threading, the I/O functions must release the GIL during the entire execution. For thread safety, they must be stateless, and the data exchanged between Python and underlying implementation should be as simple as possible. It is also important to minimize memory copies. For these reasons, we implement media functions from scratch, ensuring that functions can release the GIL and they do not introduce unnecessary overhead.
\subsubsection{GPU transfer functions}
    We also implemented functions that transfer data to GPUs, allowing SPDL to not depend on particular deep-learning library. Thus, it can be used with PyTorch, TensorFlow \cite{tensorflow}, JAX \cite{jax}, \cite{numba} etc. 
\subsection{Support for process pool}
    In practice, there are cases where functions from third-party libraries do not release the GIL.
    In such cases, it is important that the data loading pipeline can utilize subprocess.
    AsyncIO \cite{asyncio} allows users to seamlessly integrate a process pool by swapping the thread pool.
    SPDL leverages this functionality and allow user code to specify the executor, so that a part of the data loading pipeline is executed in subprocesses without being constrained by the GIL.

\subsection{Examples}
\subsubsection{Pipeline Construction}

\begin{listing}[t]
\label{lst:pipeline}
\caption{An example to build a Pipeline. The pipeline is composed of four stages. These stages are defined by the user-defined functions of which signatures are outlined at the top. The function can be a coroutine function. Each stage can configure its concurrency separately.}
\begin{verbatim}
from spdl.dataloader import PipelineBuilder

def source() -> Iterable[str]:
  # Generate series of URL for data
  ...

async def download(url: str) -> bytes:
  # Download the data from the given URL
  ...

def decode(data: bytes) -> ImageFrames:
  # Decode the downloaded data
  ...

def batch_transfer(
  frames: list[ImageFrames]) -> Tensor:
  # Create a batch from decoded data and
  # transfer to GPU

pipeline = (
  PipelineBuilder()
  .add_source(source())
  .pipe(download, concurrency=12)
  .pipe(decode, concurrency=4)
  .aggregate(32)
  .pipe(batch_transfer)
  .add_sink(buffer_size=3)
  .build(num_threads=16)
)

with pipeline.auto_stop():
  for batch in pipeline:
    ...
\end{verbatim}
\vskip -0.1in
\end{listing}

Listing \ref{lst:pipeline} shows how to build data processing pipelines from functions. The \il{PipelineBuilder} class is an interface for building a pipeline. The pipeline starts from a source object added with \il{add\_source} method. Any object with \il{Iterable} or \il{AsyncIterable} interface can be used as a source object. Usually, a source object generates a series of source locations, such as URL or file path.
After the first stage is defined, the remaining stages can be added with \il{pipe} or \il{aggregate} methods. \il{pipe} is used to chain the series of functions for processing data. Typically, functions for data acquisition, decoding media, pre-processing and device transfer are chained. Functions passed to \il{pipe} can be synchronous or asynchronous. Synchronous functions are converted to asynchronous functions internally, using thread pool. \il{pipe} also takes \il{concurrency} parameter which  controls at most how many tasks should be created using the given function. One can specify different \il{concurrency} values for different stages. For example, one can choose different concurrency for downloading and media decoding. Such granularity is necessary when tuning the performance of a pipeline. The
\il{aggregate} method can be used when it is desirable to perform a task in batch. For example, downloading data from remote storage can be faster when multiple items are requested in a single network call, or when creating a batch object, aggregating the intermediate data are necessary. Once the data processing is added, then calling \il{add\_sink} method will add a buffer where the resulting data are stored while waiting to be picked up by the main thread.
Finally, the \il{build} method creates the actual pipeline object. The size of the thread pool used by the event loop can be specified here. The resulting object is a \il{Pipeline} class. This class implements the \il{Iterable} protocol, so fetching data from the pipeline is as simple as a \il{for}-loop. Since the \il{Pipeline} uses multi-threading, the pipeline resource must be cleaned up properly before the program exists. Unlike sub-processes, there is no way to abruptly kill sub-threads, and leaving an active, non-daemonic thread can interrupt the termination of the Python interpreter indefinitely. To facilitate the clean-up, \il{Pipeline.auto\_stop} method gives a context manager, which stops and destroy background threads automatically when exiting.

\subsubsection{I/O Functions}

\begin{listing}[t]
\label{lst:io_example}
\caption{
Examples of SPDL I/O functions.
All the functions release the GIL as soon as they are called.
Decoded images are not converted to Tensor automatically.
Instead they are copied to pre-allocated, page-locked memory directly.
This minimizes the memory copy when making a batch in GPU.
}
\begin{verbatim}
import spdl.io

# Decode image w/o converting to Tensor
def decode(data: bytes) -> ImageFrames:
  packets = spdl.io.demux_image(src)
  return spdl.io.decode_packets(packets)

# pre-allocate page-locked memory and
# a CUDA stream for 
# background data transfer
STORAGE = spdl.io.cpu_storage(
  size, pin_memory=True)
STREAM = torch.cuda.Stream()

# Batch and transfer decoded images
def batch_transfer(
    frames: list[ImageFrames]) -> Tensor:
  # Copy images to pre-allocated buffer
  buffer = spdl.io.convert_frames(
    frames, storage=STORAGE)
  # Transfer the data in a non-default
  # stream to not interrupt the model
  cuda_buffer = spdl.io.transfer_buffer(
    cpu_buffer,
    device_config=spdl.io.cuda_config(
      device_index=0,
      stream=STREAM.cuda_stream,
    )
  )
  # Cast to tensor without copy
  return spdl.io.to_torch(cuda_buffer)

\end{verbatim}
\vskip -0.25in
\end{listing}

Listing \ref{lst:io_example} shows how to use SPDL I/O functions.
All the functions in SPDL I/O are implemented in a way that releases the GIL as soon as the execution moves to the low-level implementation in C++.

Since we want to minimize the memory copy, SPDL's I/O functions do not return an array.
Instead, they return a custom structure that holds the data processed by the underlying library, FFmpeg \cite{ffmpeg}.
When multimedia data are decoded by FFmpeg, they are stored as series of smaller memory chunks.
The decoding is performed by \il{demux\_image} and \il{decode\_packets}.
To convert it to an array format, we need to allocate a block of contiguous memory, and copy data.
However, since our goal is to create a batch of samples, allocating contiguous memory for each sample is wasteful therefore a block contiguous memory is created when creating a batch for the first time. \il{convert\_frames} function does this in the example.
Optionally \il{convert\_frames} takes pre-allocated and possibly page-locked memory, so that the batch is created in page-locked memory directly. The use of page-locked memory allows SPDL to overlap the data transfer and model computation.
The resulting buffer structure supports the array interface protocol, so it can be converted to formats like NumPy Array or PyTorch Tensor without creating a copy.

\section{Benchmark} \label{section:benchmark}

In this section, we measure the performance of SPDL and compare it against other solutions.
We start by measuring the throughput of the data loader without downstream load, then we add model inference (forward pass) and training (backward pass and optimizer step).
We compare the performance against publicly available solutions such as PyTorch DataLoader and NVIDIA DALI\footnotemark{}.
We use ImageNet dataset and ViT-B/16 model implementation from TorchVision \cite{torchvision2016}.
All the benchmarks were performed on AWS \il{p4d.24xlarge} instance, which has 96 CPU cores and 8 A100 GPUs. We only used one GPU. We used Python 3.12, PyTorch 2.5, (CUDA 12.1), Pillow 10.2.0 and DALI 1.42.0.
Due to space constraints we describe other benchmarks like video decoding in Appendix \ref{appendix:video_decoding}. 

\footnotetext{DALI supports CPU-only image processing and CPU/GPU mixed decoding.
The mixed decoding is faster on lower concurrency count, but on higher concurrency count, they showed similar performance, so we show the results for the mixed decoding.}

\subsection{Data Loading without Downstream Load}

\begin{figure}[t]
\begin{center}
\caption{The throughput of data loaders without downstream load.}
\vskip -0.1in
\includegraphics{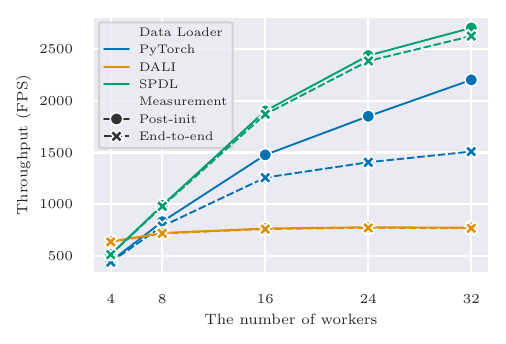}
\label{fig:benchmark_dataloading}
\end{center}
\vskip -0.3in
\end{figure}

First, we examined the throughput of data loading pipelines by themselves.
The pipelines load images from a local file system, resize, batch and transfer them to a GPU.
Fig. \ref{fig:benchmark_dataloading} shows the result.

The performance of SPDL and PyTorch improves as the number of workers is increased. DALI's performance also improves but modestly compared to the other solutions.

PyTorch DataLoader takes a non-trivial amount of time to initialize.
TorchVision's ImageNet dataset first materializes the list of paths, then PyTorch DataLoader copies them to all the workers.\footnotemark{}
Therefore the throughput of PyTorch DataLoader is very different whether such initialization is included or not.
Table. \ref{tbl:ttfb} shows the time it takes for PyTorch DataLoader to return the first batch, which is an approximation of the time it took to initialize.
Even when we exclude the initialization time, SPDL is still faster than PyTorch.

\begin{table}[t]
\caption{The time it takes for data loading pipeline to make the first batch available. Unit [second]}
\begin{center}
\begin{small}
\begin{tabular}{c r r r r r}
\hline
Concurrency &    4 &    8 &    16 &    24 &    32 \\
\hline
PyTorch     & 58.7 & 91.4 & 141.4 & 208.4 & 277.3 \\
\hline
\end{tabular}
\label{tbl:ttfb}
\end{small}
\end{center}
\vskip -0.3in
\end{table}

\footnotetext{Such copy can be easily avoided; however, generally speaking, there is no systematic way to prevent it, and we use the official implementations as-is.}

\begin{figure}[t]
\begin{center}
\caption{The CPU utilization without downstream load. The utilization was sampled where the pipeline is stable. On average, SPDL's total CPU utilization is 38\% lower than that of PyTorch DataLoader.}
\vskip -0.1in
\includegraphics{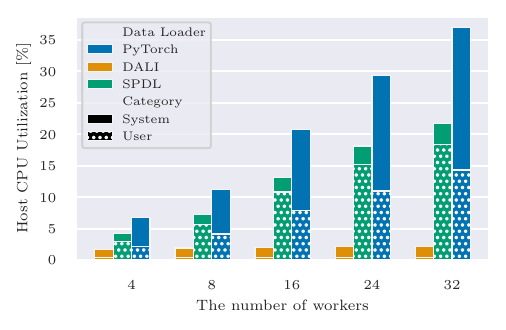}
\label{fig:benchmark_dataloading_cpu}
\end{center}
\vskip -0.3in
\end{figure}

Fig. \ref{fig:benchmark_dataloading_cpu} shows the CPU utilization at where the pipeline throughput is stable.
The CPU utilization of both PyTorch and SPDL increase as the number of workers is increased.
DALI's CPU utilization stays constant, which correlates with its throughput.
SPDL's CPU usage is mostly in user space, but the majority of PyTorch's CPU usage is in system space.
This suggests that SPDL is spending most of its CPU cycles for data loading operations, whereas PyTorch spends more than half of its CPU cycles in inter-process communication.

\begin{figure}[t]
\caption{The memory utilization of data loaders.}
\vskip -0.1in
\begin{center}
\includegraphics{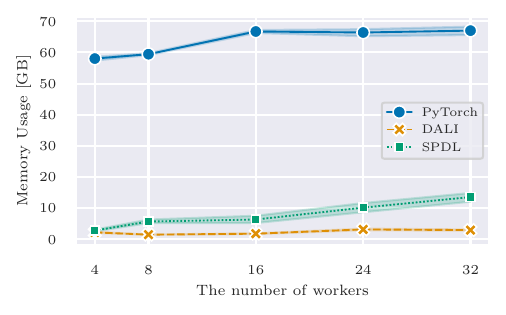}
\label{fig:benchmark_dataloading_memory}
\end{center}
\vskip -0.3in
\end{figure}

Fig. \ref{fig:benchmark_dataloading_memory} shows the memory utilization over the course of data loading.
It is notable that the PyTorch DataLoader accumulates an excessive amount of memory.
We verified that about 15 GB of it is from the list of file paths duplicated in each process workers, and the rest is accumulated over the course of data loading.

\subsection{Data Loading with Model Inference} \label{section:benchmark_inference}

\begin{figure}[t]
\caption{The end-to-end throughput of image classification inference with ViT model, using PyTorch DataLoader, Dali, and SPDL.}
\vskip -0.3in
\begin{center}
\includegraphics{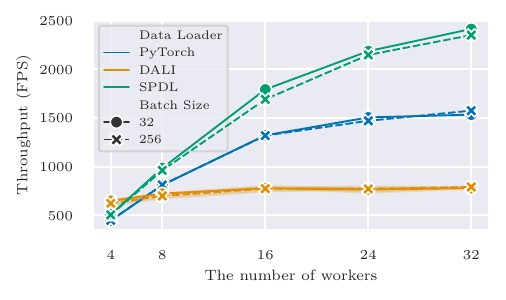}
\label{fig:benchmark_imagenet_classification}
\end{center}
\vskip -0.7in
\end{figure}

Next, we added model inference to the pipeline.
We used ViT-B/16 model implementation from TorchVision.
We cast the model to \il{bfloat16} and applied \il{torch.compile} with \textit{"max-autotune"} configuration.
Compiling the model fuses the CUDA kernels and enables CUDA Graphs \cite{compile}.
Running the resulting model launches only a handful of CUDA kernels.
For PyTorch DataLoader, the batch is transferred to the GPU in the main process then successively fed to model.
SPDL and DALI transfer the batch to the GPU in the background.

Fig. \ref{fig:benchmark_imagenet_classification} shows the end-to-end throughput of each pipeline.
Overall the performance trend is similar to that of Fig. \ref{fig:benchmark_dataloading}.
We changed the batch size from 32 to 256, but it did not have a significant effect on throughput.
The CPU and memory utilization are similar to those without the inference load. They are found in Appendix \ref{appendix:system_stas}

\subsection{Data Loading with Model Training} \label{section:benchmark_training}

\begin{figure}[t]
\begin{center}
\caption{The end-to-end throughput of ViT model training on ImageNet dataset. SGD optimizer was used. The \il{MAX} indicates the maximum throughput obtained by running the model without a data loader.}
\vskip -0.15in
\includegraphics{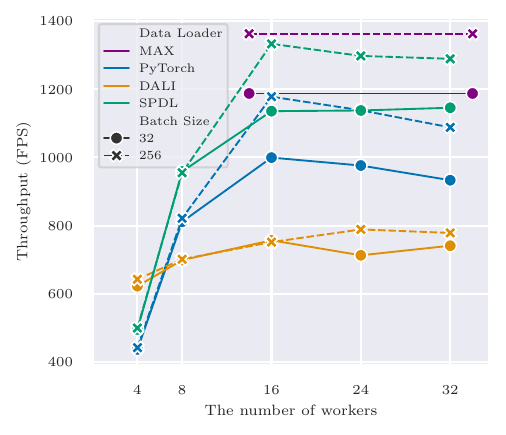}
\label{fig:benchmark_imagenet_training}
\end{center}
\vskip -0.6in
\end{figure}

Finally, we measured the performance of a training pipeline. We added SGD optimizer step to the pipelines.
We also added a dummy data loader to see the maximum achievable throughput.
The dummy data loader creates a fake batch tensor once and returns it on every iteration, so it does not add any delay to the data loading or load to the system.
Fig. \ref{fig:benchmark_imagenet_training} shows the throughput of the pipelines.
SPDL consistently outperforms PyTorch DataLoader, and its peak performance is very close to the maximum throughput from the dummy data loader.
The CPU and memory utilization are found in Appendix \ref{appendix:system_stas}

\section{Free-Threaded Python} \label{section:ft}

\begin{figure}[t]
\caption{The throughput of SPDL over different Python versions including FT Python and PyTorch DataLoader as a baseline.}
\includegraphics{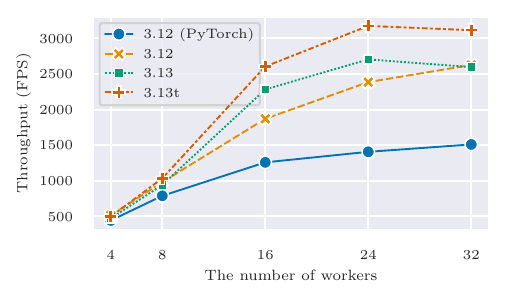}
\label{fig:benchmark_ft_scalability}
\end{figure}

\begin{table}[t]
\caption{The peak throughput of SPDL (without downstream load) over different Python versions and PyTorch as reference.}
\label{table:peak}
\begin{center}
\begin{small}
\begin{tabular}{c c c c | c }
\hline
Solution & 3.13t & 3.13 & 3.12 & 3.12 (PyTorch) \\
\hline
Throughput (FPS) & 3158  & 2714 & 2627 & 1510 \\
\hline
\end{tabular}
\end{small}
\end{center}
\vskip -0.2in
\end{table}

Lastly, to get the answer to the original questions, ("How much speed up can we get from FT-Python?" and "Is it possible to bring the benefit to the current Python?"), we measured the performance of SPDL on Python 3.13 and 3.13t.

The Fig. \ref{fig:benchmark_ft_scalability} shows the throughput.
The throughput increases as we switch the data loader from PyTorch to SPDL, then Python version 3.12 to 3.13 and finally to Free-Threaded Python.
The Table. \ref{table:peak} summarizes the peak throughput. When compared against PyTorch DataLoader on 3.12, SPDL on 3.13t is x2.1 faster, and SPDL on 3.12 has already realized 67\% of such speed up. 

It is worth noting that the source code for this benchmark is not changed when running on 3.13t.

\section{Conclusion}

In this paper, we present SPDL, a framework-agnostic library for data loading.
SPDL is resource efficient because it embraces thread-based parallelism, and does not suffer from IPC overhead.
As a result SPDL is faster than process-based solutions like PyTorch DataLoader, despite it operating under the constraint of the GIL in Python versions 3.12 and lower.
SPDL's design is also compatible with Free-Threaded Python (Python 3.13t) and provides additional performance boosts without modifying any code.
SPDL paves a smooth path to adopt multi-threading for when FT-Python becomes production-ready.

\newpage

\bibliography{mlsys2025/citation}
\bibliographystyle{mlsys2025/mlsys2024}


\newpage
~
\newpage

\appendix

\section{CPU and memory utilization in ImageNet Inference} \label{appendix:system_stas}

\begin{figure}[hb]
\caption{The CPU utilization of PyTorch, SPDL and DALI when running inference with ViT b/16 model on ImageNet train split.}
\vskip -0.3in
\begin{center}
\includegraphics{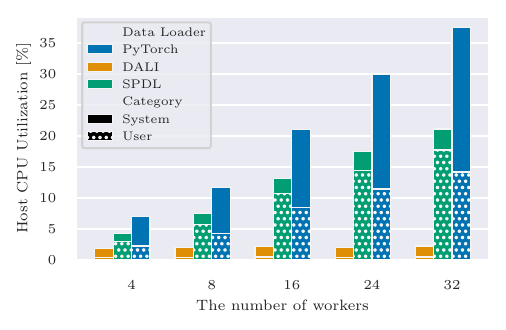}
\label{fig:benchmark_imagenet_classification_cpu}
\end{center}
\vskip -0.7in
\end{figure}

\begin{figure}[hb]
\caption{The memory utilization of PyTorch, SPDL and DALI when running inference with ViT b/16 model on ImageNet train split.}
\vskip -0.3in
\begin{center}
\includegraphics{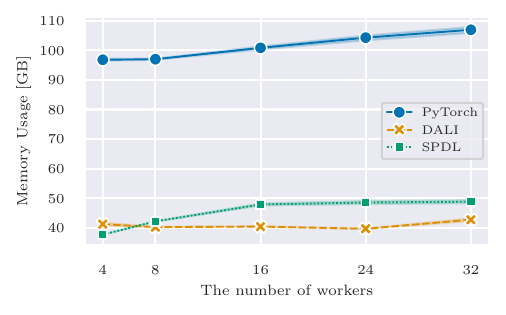}
\label{fig:benchmark_imagenet_classification_memory}
\end{center}
\vskip -0.7in
\end{figure}

Fig \ref{fig:benchmark_imagenet_classification_cpu} and \ref{fig:benchmark_imagenet_classification_memory} show the CPU and memory utilization of each pipeline from Section \ref{section:benchmark_inference}

\begin{figure}[t]
\caption{The CPU utilization of PyTorch, SPDL and DALI when training ViT b/16 model with ImageNet train split.}
\vskip -0.3in
\begin{center}
\includegraphics{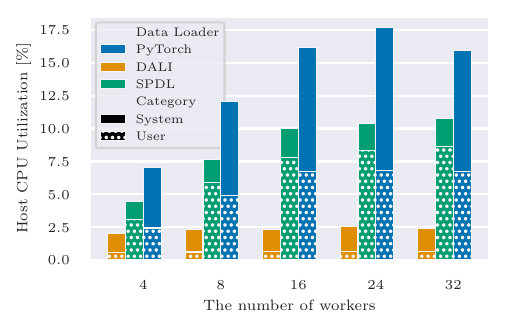}
\label{fig:benchmark_imagenet_training_cpu}
\end{center}
\vskip -0.7in
\end{figure}

\begin{figure}[t]
\caption{The memory utilization of PyTorch, SPDL and DALI when training ViT b/16 model with ImageNet train split.}
\vskip -0.3in
\begin{center}
\includegraphics{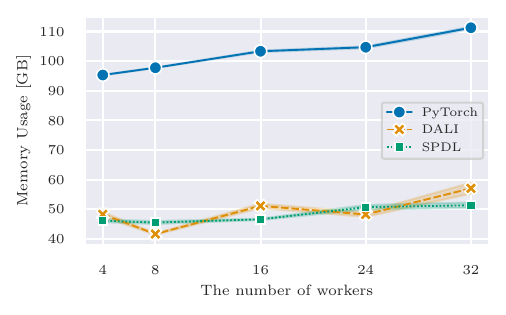}
\label{fig:benchmark_imagenet_training_memory}
\end{center}
\vskip -0.7in
\end{figure}

Fig \ref{fig:benchmark_imagenet_training_cpu} and \ref{fig:benchmark_imagenet_training_memory} show the CPU and memory utilization of each pipeline from Section \ref{section:benchmark_training}.

\newpage
~
\newpage

\section{Performance comparison of multi-processing and multi-threading} \label{appendix:process_vs_thread}

Listing. \ref{lst:process_vs_thread} is the core implementation used to obtain Fig. \ref{fig:process_vs_thread} from Section \ref{section:process_vs_thread} to measure the performance of Pillow. 

It uses the ImageNet dataset from TorchVision library, which wraps the Pillow operations. The function is executed in \il{ThreadPoolExecutor}. Batch tensors are created by collating mulitple tensors. This is usually handled by PyTorch's \il{DataLoader} class, but here, we use custom implementation and call the collate function directly. This allows simple comparison between multi-threading and multi-processing by swapping \il{ThreadPoolExecutor} and \il{ProcessPoolExecutor}.

We measured the time since the first job is submitted till the last job is completed\footnotemark{}, while changing the concurrency and batch size. As we saw in Table. \ref{tbl:ttfb}, the time to initialize the process workers is non-trivial, but here we focus on the post-initialization performance.

Listing. \ref{lst:process_vs_thread_spdl} shows how to swap the Pillow with SPDL's IO module.
The same implementation was used to measure the throughput on Python 3.12 and 3.13t.

\footnotetext{The thread/process pool executors are initialized before the timer starts. The dataset instance is created in each process beforehand. Only sample indices and resulting batch are exchanged between processes.}

\begin{listing}[h]
\label{lst:process_vs_thread}
\caption{Code for measuring the performance of data loading. By swapping \il{ThreadPoolExecutor} with \il{ProcessPoolExecutor}, we can compare the performance of thread-based parallelism and process-based parallelism.}
\begin{verbatim}

from concurrent.futures import \
    ThreadPoolExecutor
from torch import utils
import torchvision.datasets
import torchvision.transforms as T

DATASET = torchvision.datasets.ImageNet(
  root_dir,
  transform=T.Compose([
    T.Resize((224, 224)),
    T.PILToTensor(),
  ]),
)

def process(indices):
  items = [DATASET[i] for i in indices]
  return utils.data.default_collate(items)

def benchmark(
  sampler: Iterable[list[int]],
  workers: int):
  with ThreadPoolExecutor(workers) as ex:
    fs = [
      ex.submit(process, indices)
      for indices in sampler]
    for f in fs:
      f.result()
\end{verbatim}
\end{listing}

\begin{listing}[h]
\label{lst:process_vs_thread_spdl}
\caption{Code for measuring the performance of data loading. By swapping \il{ThreadPoolExecutor} with \il{ProcessPoolExecutor}, we can compare the performance of thread-based parallelism and process-based parallelism.}

\begin{verbatim}
from spdl.io import (
  ImageFrames,
  get_video_filter_desc,
)

def decode_image(path) -> ImageFrames:
  filter_desc = get_video_filter_desc(
    scale_width=244,
    scale_height=244,
    pix_fmt="rgb24",
  )
  packets = spdl.io.demux_image(path)
  frames = spdl.io.decode_packets(
    packets, filter_desc=filter_desc)
  buffer = spdl.io.convert_frames(frames)
  return spdl.io.to_torch(buffer)

DATASET = ImageNet(
    root_dir,
    transform=None,
    loader=default_loader,
)
\end{verbatim}
\end{listing}

\section{Benchmarking on Video Decoding} \label{appendix:video_decoding}

For the video processing, we compare the solution against Decord.

Comparing the performance fairly against Decord is not strightforward because the resource utilization of Decord is unbounded as described in \ref{section:decord}. More specifically, Decord's \il{VideoLoader} opens all the videos from the given list at initialization, then start decoding them in the background. When decoding videos it lets FFmpeg chooses the optimal number of threads that should be used for decoding a single video. These decoder threads are initialized and destroyed for each video file.

The table \ref{tbl:decord_ttfb} shows the time it takes for decord's \il{VideoLoader} to initialize the \il{VideoLoader} class and returns the first batch.

\begin{figure}[t]
\caption{Performance comparison between SPDL and Decord. Note that the values of decord are placed there just because SPDL gives similar throughput at 4 workers. Decord does not give control over the thread concurrency. It processes videos aggressively in a backgroun d thread, while instructing FFmpeg to use multiple threads for decoding, which makes it difficult to quantify the concurrency.}
\includegraphics{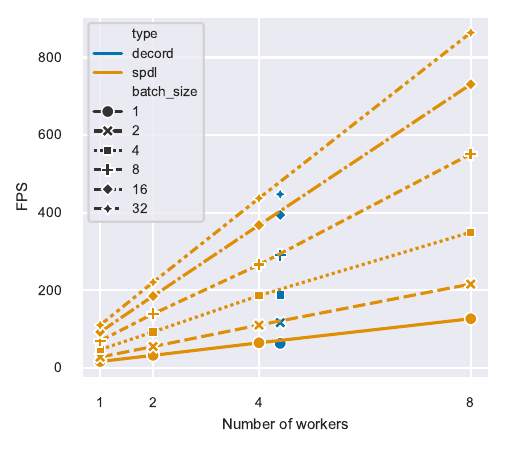}
\label{fig:spdl_vs_decord}
\vskip -0.25in
\end{figure}

The Fig. \ref{fig:spdl_vs_decord} shows the throughput of SPDL and Decord on loading videos and sampling frames. A list of 10000 videos from Kinetic 400 \cite{kinetics} was given, and batches were extracted with 5 intra-batch frames and 50 inter-batch frames. It measures the time between the start of initialization and at most 1000 batches are returned.

Decord performs all the video processing in C++, without being constrained by the GIL, while SPDL uses Python threads, potentially blocked by the GIL. SPDL can achieve similar performance as Decord using a few threads. It is worth noting that SPDL is robust against decoding failure, while Decode fails if any of video processing fails.
\begin{table}[h]
\caption{The initialization time for decord's \il{VideoLoader}. It sequentially processes all the videos at initialization, so it takes more time to initializes as the number of videos increases.}
\vskip 0.5em
\label{tbl:decord_ttfb}
\begin{center}
\begin{small}
\begin{tabular}{c c c c c c}
\hline
Number of videos & 1000 & 2000 & 4000 & 8000  & 10000 \\ \hline
Time {[}sec{]}   & 6.3  & 13.0 & 27.4 & 53.25 & 67.5  \\
\hline
\end{tabular}
\end{small}
\end{center}
\vskip -0.1in
\end{table}

\end{document}